\documentclass{rmf-d}
\usepackage{nopageno,rmfbib,multicol,times,epsf,amsmath,amssymb,cite}
\usepackage[latin1]{inputenc}
\usepackage[]{caption2}
\usepackage{graphics}

\usepackage{amsmath}
\usepackage[usenames]{color}
\usepackage{epsfig}
\usepackage{epstopdf}
\usepackage{amssymb,amsmath}
\usepackage{amsmath}
\usepackage{epstopdf}
\usepackage{sidecap}
\usepackage{cases}
\usepackage{enumitem}
\usepackage{bm}
\usepackage{overpic}
\usepackage{upgreek}
\usepackage{morefloats}

\usepackage{hyperref}
\def\rmfcornisa{Research OR Education of Physics \hfill\rmf\ {\bf ??} (*?*) ???--???
\hfill MES? A\~NO?}

\newcommand{\HISKP}{Helmholtz-Institut f\"{u}r Strahlen- und Kernphysik, Universit\"{a}t Bonn, Germany}

\newcommand{\Ocal}{\mathcal{O}}

\def\rmfcintilla{{\it Rev.\ Mex.\ Fis.\/} {\bf ??} (*?*) (????) ???--???}
\clearpage \rmfcaptionstyle 
\begin{document}
\title{  Determination of complete experiments using graphs
\vspace{-6pt}}
\author{ Y.~Wunderlich, P.~Kroenert, F.~Afzal and A.~Thiel  }
\address{ \HISKP }
\author{ }
\address{ }
\author{ }
\address{ }
\author{ }
\address{ }
\author{ }
\address{ }
\author{ }
\address{ }
\maketitle
\recibido{day month year}{day month year
\vspace{-12pt}}
\begin{abstract}
\vspace{1em} 
This work presents ideas for the determination of complete experiments using graphs, which are based on a recently published, modified form of Moravcsik's theorem. The lucid representation of complete experiments in terms of graphs, which is at the heart of the theorem, leads to a fully automated procedure that can determine complete experiments for in principle any reaction, i.e. for any number of amplitudes~$N$. For larger~$N$ (i.e.~$N \geq 4$), the sets determined according to Moravcsik's theorem turn out to be slightly overcomplete. A new type of directional graph has been proposed recently, which can decrease the length of the complete sets of observables in some of these cases. The presented results are relevant for reactions with larger numbers of spin-amplitudes, which are at the center of interest in forthcoming measurements, such as single-meson electroproduction~$(N=6)$, two-meson photoproduction~$(N=8)$ or vector-meson photoproduction~$(N=12)$.   
\vspace{1em}
\end{abstract}
\keys{  hadron spectroscopy, baryon spectroscopy, photonuclear reactions, amplitude-extraction problems, polarization observables, complete experiments, Moravcsik graphs  \vspace{-4pt}}
\pacs{ 02.30.Zz, 13.88.+e, 13.60.Le, 14.20.Gk, 24.70.+s, 25.20.Lj    \vspace{-4pt}}
\begin{multicols}{2}

\section{Introduction} \label{sec:Intro}

The spectroscopy of baryons~\cite{Klempt:2009pi,Crede:2013kia,Ireland:2019uwn} is of vital importance for our understanding of strong~QCD and thus many experimental programs are still ongoing at facilities around the world~\cite{Thiel:2012yj,Gottschall:2013uha,Hartmann:2014mya,CBELSA/TAPS:2020yam,Dugger:2013,Collins:2017sgu,Senderovich:2015lek,Strauch:2015zob,Burkert:2019kxy,Burkert:2020akg,Hornidge:2012ca,Adlarson:2015byy,Gardner:2016irh,Annand:2016ppc,Kashevarov:2017kqb,Briscoe:2019cyo,Kohri:2017kto,Kohri:2020ucd,AlGhoul:2017nbp,Adhikari:2019gfa,Adhikari:2020cvz}. Reactions among particles with spin are of interest in baryon spectroscopy, due to the fact that resonant intermediate states have half-integer spin~$J$. For such reactions, one encounters the ubiquitous amplitude-extraction problems, where a set of $N$ spin-amplitudes (either helicity- or transversity amplitudes) has to be extracted from $N^{2}$ measurable polarization observables. These problems naturally motivate the search for so-called {\it complete experiments} (or {\it complete sets of observables})~\cite{Barker:1975bp,Keaton:1995pw,Chiang:1996em,Nakayama:2018yzw,Wunderlich:2020umg,Wunderlich:2021xhp}, which are (possibly minimal) subsets selected from the full set of $N^2$ observables, that allow for a unique extraction of the amplitudes. Convincing heuristic arguments show that the absolutely minimal length of a complete experiment is~$2 N$ (see for instance ref.~\cite{Keaton:1996pe} or the introductions of references~\cite{Wunderlich:2020umg,Wunderlich:2021xhp}), for arbitrary~$N$. This work discusses methods for the selection of complete experiments, which all use the language of graphs.

\section{Amplitude-extraction problems} \label{sec:CompExps}

For a generic amplitude-extraction problem with $N$ transversity amplitudes $\left\{ b_{i}, i = 1,\ldots,N \right\}$, the corresponding set of $N^{2}$ (polarization-) observables is given as~\cite{Wunderlich:2020umg}
\begin{equation}
 \Ocal^{\alpha} = \bm{c}^{\alpha} \sum_{i,j = 1}^{N} b_{i}^{\ast} \tilde{\Gamma}^{\alpha}_{ij} b_{j}  , \text{ for } \alpha = 1,\ldots,N^{2}, \label{eq:GenericObservableDef}
\end{equation}
where~$\bm{c}^{\alpha}$ are (possibly) observable-dependent normalization factors and the~$\tilde{\Gamma}^{\alpha}$ represent an $N \times N$ Clifford algebra~\cite{Chiang:1996em}.

\begin{figure}[ht]
\centering
\includegraphics[width=0.825\textwidth,trim = 0 5.0cm 0 1.1cm,clip]{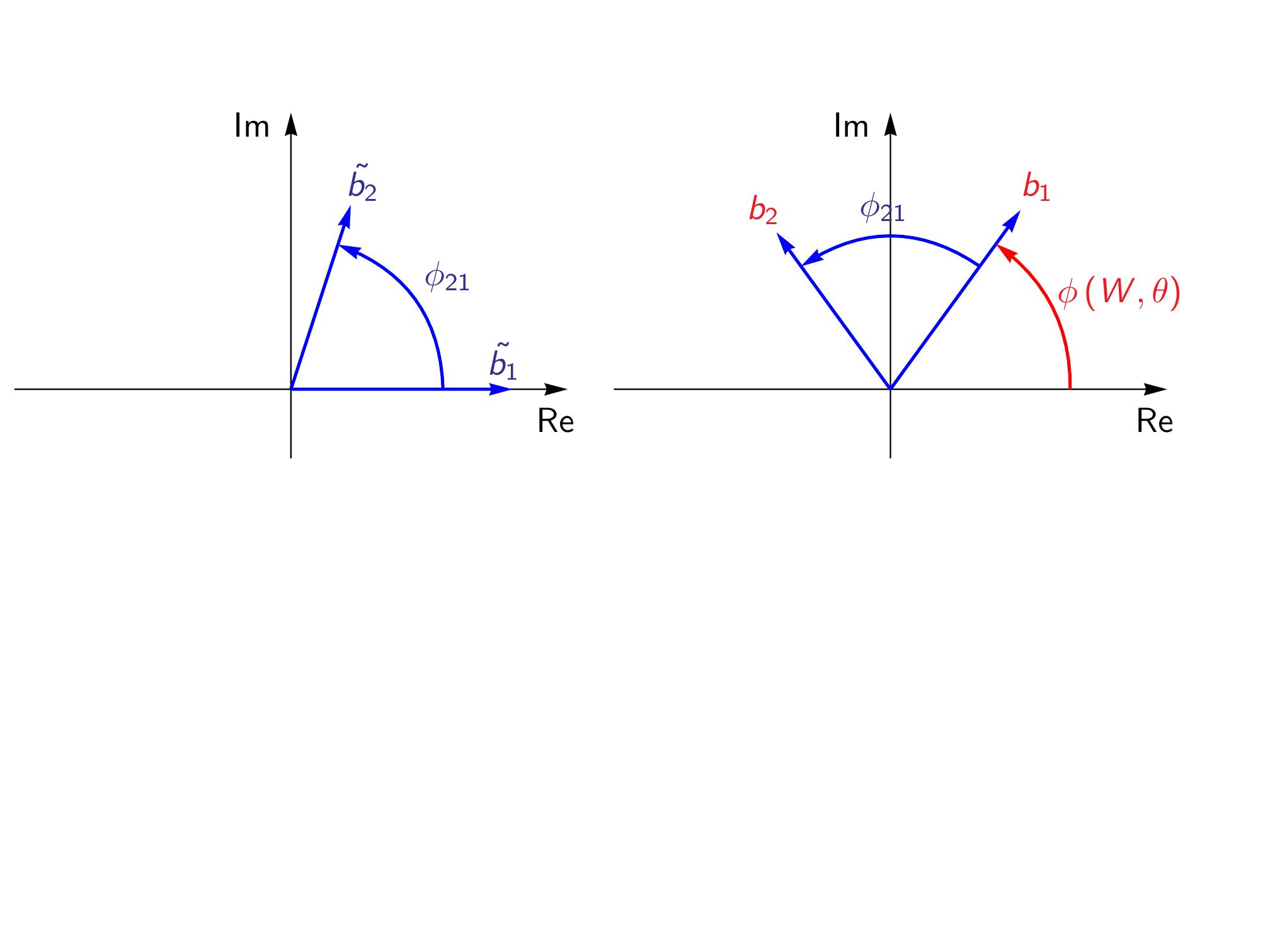} 
\caption{The basic amplitude-extraction problem defined in section~\ref{sec:CompExps} is illustrated for the simple case of~$N = 2$ amplitudes. A complete experiment can fix all the amplitudes uniquely in case a phase-convention is imposed (left), however the real solution has one unknown overall phase~$\phi (W, \theta)$ which cannot be determined by the complete experiment (right).}
\label{fig:CompExpTwoAmplitudes}
\end{figure}

The minimization of measurement effort motivates now the search for the above-mentioned \textit{complete experiments}. However, due to the bilinear nature of the equations~\eqref{eq:GenericObservableDef}, any complete set of observables can only maximally constrain the~$N$ amplitudes up to one unknown overall phase. In other words, one can only determine moduli and relative phases (Figure~\ref{fig:CompExpTwoAmplitudes} illustrates this fact for~$N = 2$). The unknown overall phase can depend on the full reaction kinematics. 

Eq.~\eqref{eq:GenericObservableDef} can be inverted in an effective linearization~\cite{Wunderlich:2020umg}:
\begin{equation}
 b_{i}^{\ast} b_{j} = \frac{1}{\tilde{N}} \sum_{\alpha = 1}^{N^{2}}  \left( \tilde{\Gamma}^{\alpha}_{ij} \right)^{\ast} \left( \frac{\Ocal^{\alpha}}{\bm{c}^{\alpha}} \right)  . \label{eq:BilProductInverted}
\end{equation}
A suitable choice for the indices~$(i,j)$ gives access to moduli and relative phases of the amplitudes. However, in this approach, the length of the complete sets of observables implied by the right-hand-side of equation~\eqref{eq:BilProductInverted} turns out to be far above the absolutely minimal number~$2 N$.

An alternative method, pioneered by Moravcsik~\cite{Moravcsik:1984uf}, is to consider directly the basis of bilinear products
\begin{equation}
 b_{j}^{\ast} b_{i}  , \text{ for } i,j=1,\ldots,N, \label{eq:GeneralBilProductBasis}
\end{equation}
and to consider combinations of real-and imaginary part of such products individually (see further details in section~\ref{sec:MoravcsiksTheorem}). 

For any algebraic solution-Ansatz, the initial standard-assumption states that the moduli $\left| b_{1} \right|, \left| b_{2} \right|, \ldots,  \left| b_{N} \right| $ are already \textit{known} from a certain subset of 'diagonal' observables. Then, a suitable minimal set of \textit{relative phases} $\phi_{ij} := \phi_{i} - \phi_{j}$ (for $b_{j} = \left| b_{j} \right| e^{i \phi_{j}}$) has to be determined. Deriving a general solution-theory for arbitrary~$N$ is very difficult in the $\Ocal^{\alpha}$-basis, but in the $b_{j}^{\ast} b_{i}$-basis a standard solution exists: Moravcsik's theorem~\cite{Moravcsik:1984uf,Wunderlich:2020umg}. In the following, we outline the road to this theorem.

\section{Ambiguities and consistency relations} \label{sec:Ambs}

\begin{figure}[ht]
\centering
\includegraphics[width=0.8\textwidth,trim = 0 5.0cm 0 1.1cm,clip]{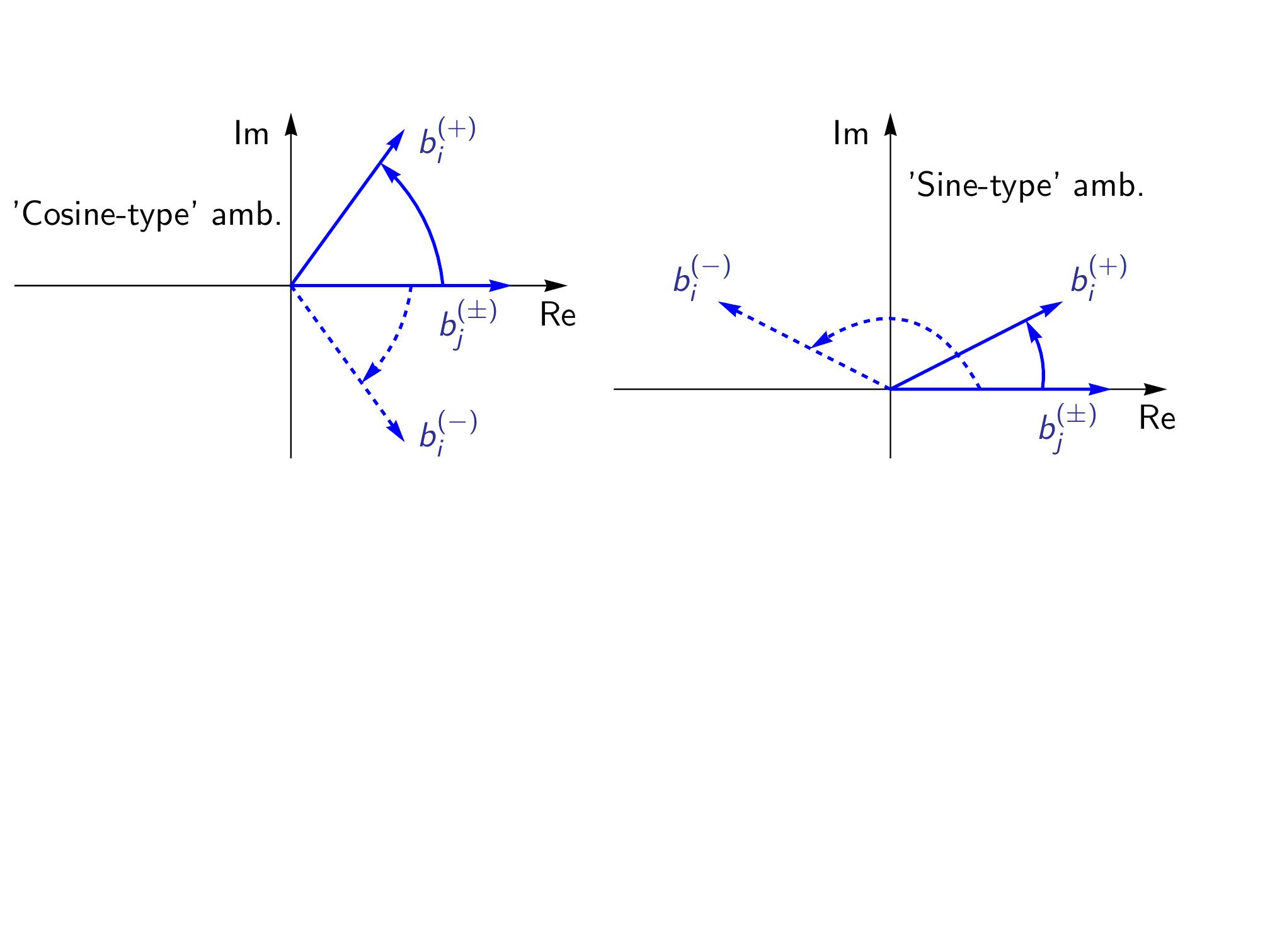} 
\caption{The basic discrete ambiguities discussed in section~\ref{sec:Ambs} are illustrated. These are the 'cosine-type' ambiguity resulting from~$\text{Re} \left[ b_{j}^{\ast} b_{i} \right]$ (left), as well as the 'sine-type' ambiguity stemming from~$\text{Im} \left[ b_{j}^{\ast} b_{i} \right]$ (right).}
\label{fig:QuadrantAmbiguities}
\end{figure}

The real part of a general bilinear product~$b_{j}^{\ast} b_{i} $ is written as
$\mathrm{Re} \left[ b_{j}^{\ast} b_{i} \right] = \left| b_{i} \right|\left| b_{j} \right|  \mathrm{Re} \left[ e^{i \phi_{ij}} \right]  = \left| b_{i} \right|\left| b_{j} \right|  \cos \phi_{ij}$.
Applying the $\arccos$ function, we see that the relative phase~$\phi_{ij}$ is fixed up to the discrete phase-ambiguity~\cite{Nakayama:2018yzw,Wunderlich:2020umg}
\begin{equation}
 \phi_{ij} \longrightarrow \phi_{ij}^{\pm} = \begin{cases}  + \alpha_{ij}, \\ - \alpha_{ij} ,   \end{cases}  \label{eq:DiscrAmbRealPart}
\end{equation}
where $\alpha_{ij} \in \left[ 0, \pi \right]$ is determined uniquely from~$\mathrm{Re} \left[ b_{j}^{\ast} b_{i} \right]$ and the known moduli. We call the discrete ambiguity resulting from the real part a 'cosine-type' ambiguity and illustrate it in Figure~\ref{fig:QuadrantAmbiguities}, on the left. Such a type of discrete ambiguity is sometimes also called a \textit{quadrant ambiguity}. 

Similarly, the imaginary part reads~$\mathrm{Im} \left[ b_{j}^{\ast} b_{i} \right] = \left| b_{i} \right|\left| b_{j} \right|  \sin \phi_{ij}$,
which implies the discrete ambiguity~\cite{Nakayama:2018yzw,Wunderlich:2020umg}
\begin{equation}
 \phi_{ij} \longrightarrow \phi_{ij}^{\pm} = \begin{cases}  + \alpha_{ij}, \\ \pi - \alpha_{ij} ,   \end{cases}  \label{eq:DiscrAmbImaginaryPart}
\end{equation} 
with a uniquely specified $\alpha_{ij} \in \left[ -\pi / 2, \pi / 2 \right]$. This ambiguity is also of the quadrant type. We call it 'sine-type' ambiguity and illustrate it in Figure~\ref{fig:QuadrantAmbiguities}, on the right.

\begin{figure}[ht]
\centering
\includegraphics[width=0.85\textwidth,trim = 0 5.0cm 0 1.1cm,clip]{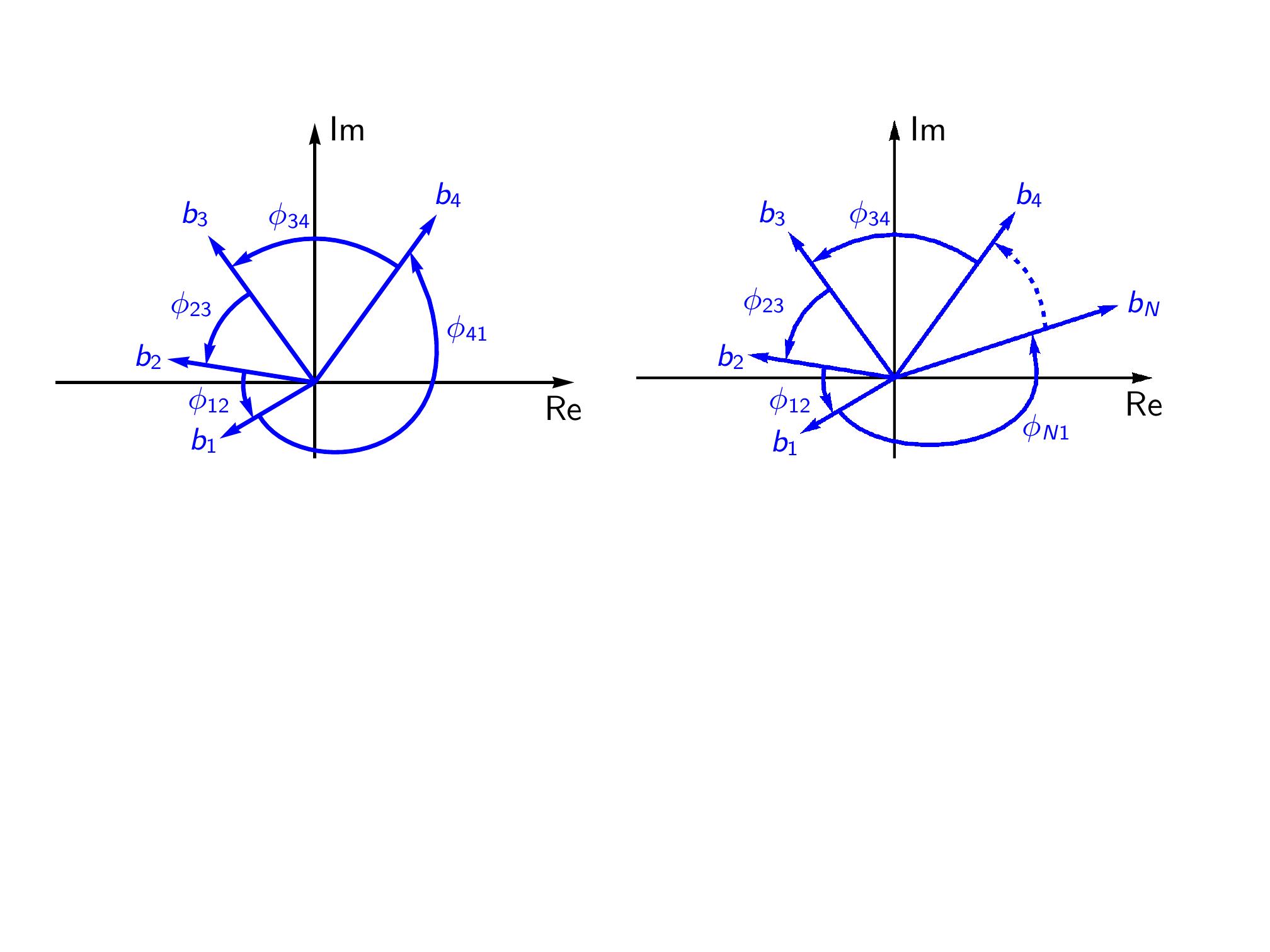} 
\caption{The consistency relations discussed in sec.~\ref{sec:Ambs} are illustrated for $N = 4$ amplitudes (left) and for the case of arbitrary $N$ (right).}
\label{fig:ConsistencyRelations}
\end{figure}

For a general problem with~$N$ amplitudes, the number of ambiguous solutions implied by equations~\eqref{eq:DiscrAmbRealPart} and~\eqref{eq:DiscrAmbImaginaryPart} grows exponentially with~$N$. The question thus arises whether one can impose additional constraints in order to reduce the number of ambiguous solutions. Such constraints can indeed be found in so-called~\textit{consistency relations}~\cite{Moravcsik:1984uf,Chiang:1996em,Nakayama:2018yzw,Wunderlich:2020umg}. The latter are quite natural constraints for the relative phases of an arrangement of~$N$ amplitudes in the complex plane and we illustrate them geometrically in Figure~\ref{fig:ConsistencyRelations}. On the left of Figure~\ref{fig:ConsistencyRelations}, we show an example for~$N = 4$ amplitudes and for this, the consistency relation corresponding to the figure reads: $\phi_{12} + \phi_{23} + \phi_{34} + \phi_{41} = 0$ (up to addition of multiples of~$2 \pi$). In the general case (cf. right-hand-side of Figure~\ref{fig:ConsistencyRelations}), a generic consistency relation reads~\cite{Moravcsik:1984uf,Wunderlich:2020umg,Wunderlich:2021xhp}
\begin{equation}
 \phi_{1i} + \phi_{ij} + \ldots + \phi_{pq} + \phi_{q1} = 0 . \label{eq:ConsistencyRelGeneralCase}
\end{equation}
%
Moravcsik's theorem~\cite{Moravcsik:1984uf,Wunderlich:2020umg}, which will be discussed in the next section, is now simply a systematic study of a combination of the discrete ambiguities~\eqref{eq:DiscrAmbRealPart} and~\eqref{eq:DiscrAmbImaginaryPart} with consistency relations. One wishes to determine when the different cases for a consistency relation implied by the ambiguities, i.e.
\begin{equation}
 \phi^{\pm}_{1i} + \phi^{\pm}_{ij} + \ldots + \phi^{\pm}_{pq} + \phi^{\pm}_{q1} = 0 , \label{eq:ConsistencyRelGeneralCase}
\end{equation}
are (linearly) independent of each other. In case full independence of all relations is achieved, the solution of the amplitude-extraction problem becomes unique.

\section{The modified form of Moravcsik's theorem} \label{sec:MoravcsiksTheorem}

The original theorem developed by Moravcsik~\cite{Moravcsik:1984uf} has been reexamined and slightly modified recently~\cite{Wunderlich:2020umg}. The theorem is formulated in the following 'geometrical analog'~\cite{Moravcsik:1984uf}:
\begin{quotation}
 We represent every amplitude~$b_{1}, \ldots, b_{N}$ by a {\it point} and every product~$b_{j}^{\ast} b_{i}$, or rel.-phase $\phi_{ij}$, by a {\it line connecting points} '$i$' and '$j$'.  Furthermore, we represent every real part $\text{Re} \left[ b_{j}^{\ast} b_{i} \right]$ $\propto$ $\cos \phi_{ij}$ by a {\it solid line} and every imaginary part $\text{Im} \left[ b_{j}^{\ast} b_{i} \right]$ $\propto$ $\sin \phi_{ij}$ by a {\it dashed line}.
\end{quotation}
The thus constructed graph is {\it fully complete} (i.e. the solution of the amplitude-extraction problem is unique) if it satisfies:
\begin{itemize}
 \item[(i)] The graph is fully {\it connected} and all points have to have {\it order two} (i.e. are attached to two lines). In this case, all continuous ambiguities are resolved~\footnote{This means that the connectedness requirement forbids combinations of relative-phases corresponding to multiple disconnected sub-sets of amplitudes in the complex plane, which can rotate freely relative to each other.} and the existence of a {\it consistency relation} is ensured. \\ The graphs required here are what Moravcsik calls the 'most economical'~\cite{Moravcsik:1984uf} possibility and thus in some sense a convenience. The number of relevant graph-topologies of this type is~$\frac{(N - 1)!}{2}$, for arbitrary~$N \geq 3$.
 \item[(ii)] The graph has to have an {\it odd} number of dashed lines, as well as {\it any} number of solid lines. In this case, all discrete ambiguities are resolved as well.
\end{itemize}
This theorem is proven in all generality in Appendix~A of reference~\cite{Wunderlich:2020umg}. In the following, we apply it to some examples.

\section{Applications: photo- and electroproduction} \label{sec:ApplicationsPhotoElectroproduction}

\begin{figure}[ht]
 \centering
  \includegraphics[width=0.2405\textwidth]{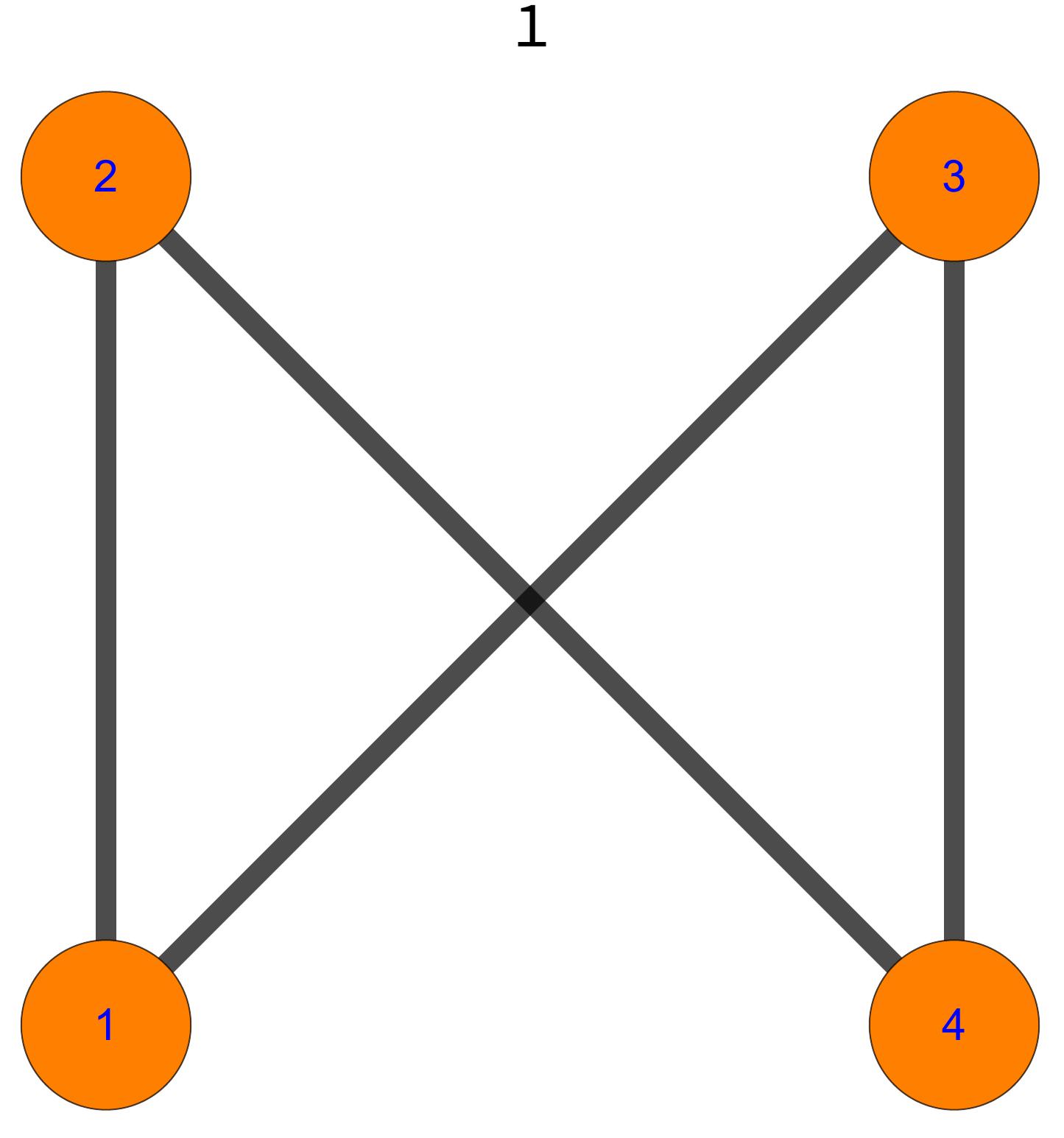}\hspace*{9.5pt}
  \includegraphics[width=0.2405\textwidth]{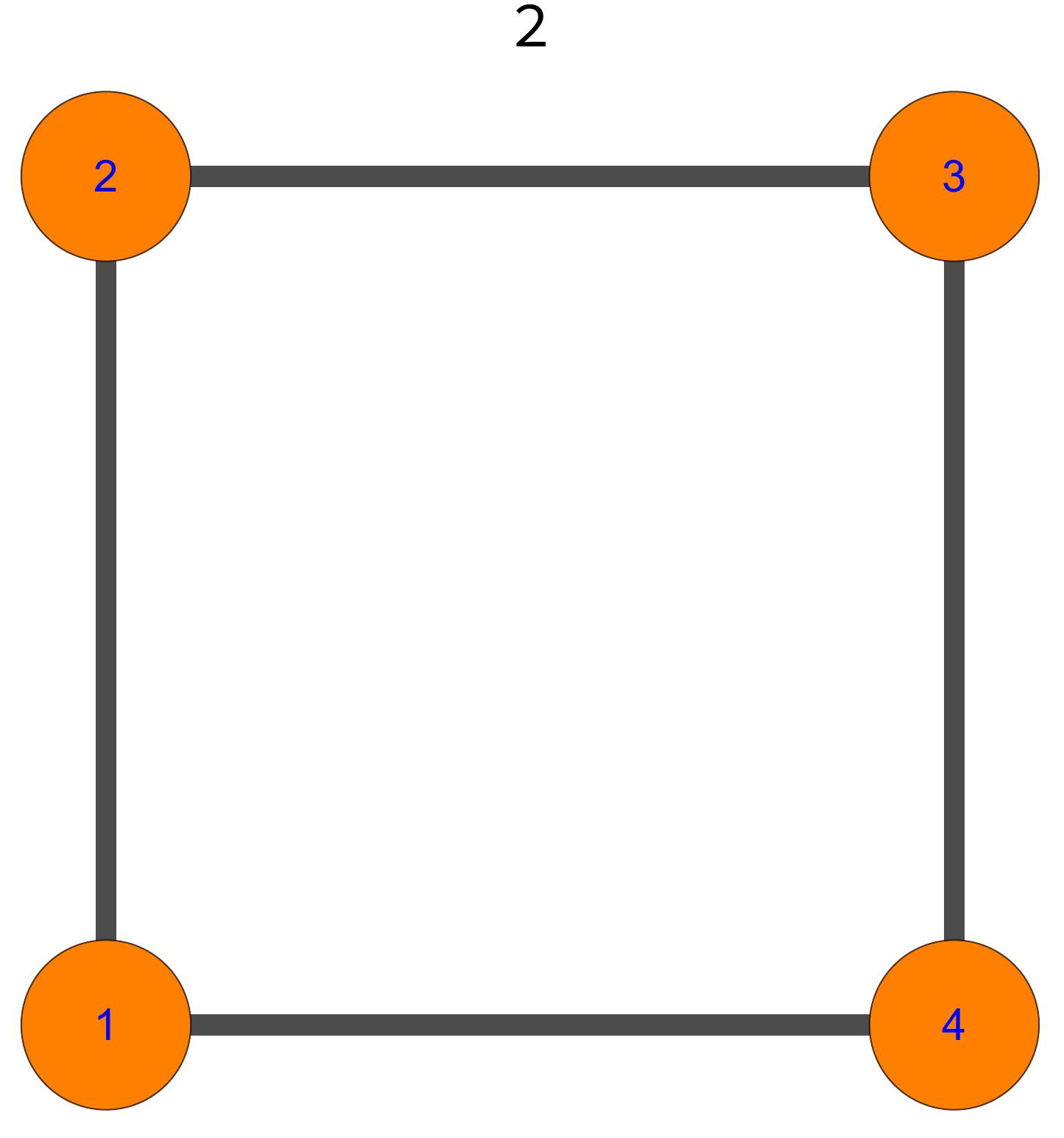}
  \hspace*{9.5pt}
  \includegraphics[width=0.2405\textwidth]{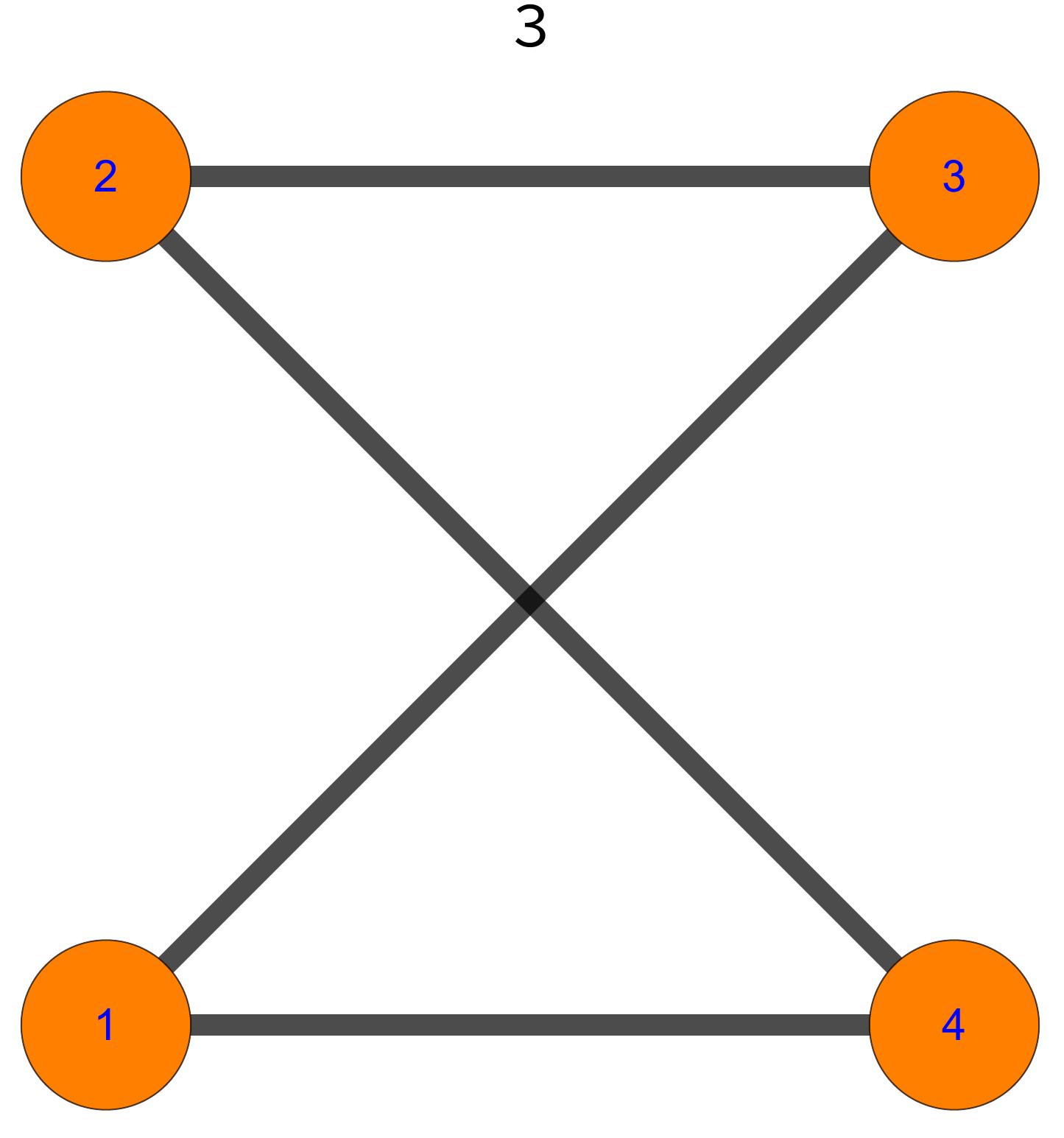}
  \caption{For photoproduction ($N = 4$ amplitudes, see section~\ref{sec:ApplicationsPhotoElectroproduction}), the three relevant start-topologies are shown (figures from ref.~\cite{Wunderlich:2020umg}).}
\label{fig:PhotoproductionStartTopologies}
\end{figure}

\begin{figure}[ht]
\centering
\includegraphics[width=0.825\textwidth,trim = 0 4.7cm 0 1.3cm,clip]{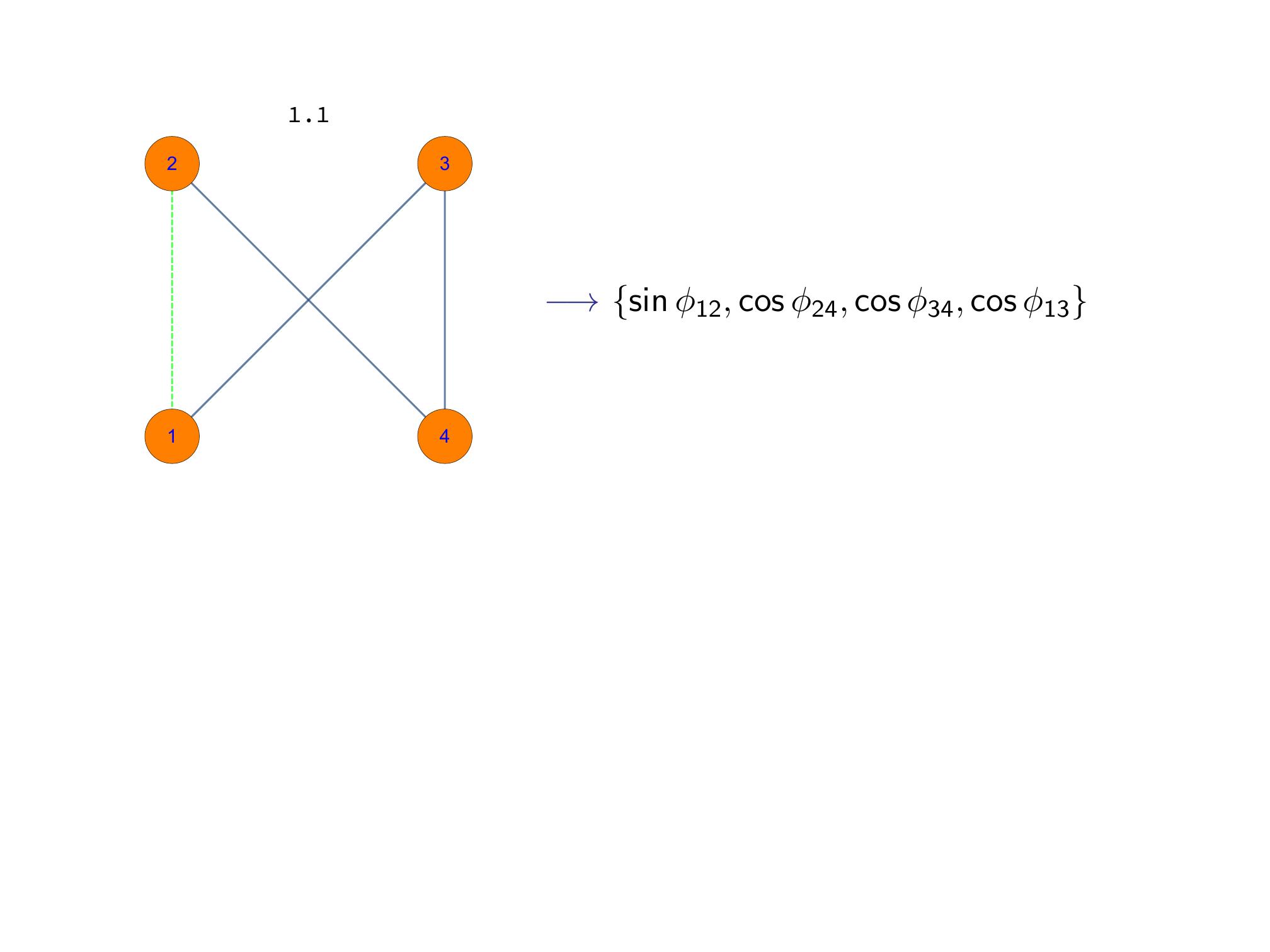} 
\vspace*{-10pt}
\caption{One example for the fully complete graphs according to the modified form of Moravcsik's theorem (cf. section~\ref{sec:MoravcsiksTheorem}), which can be derived from topology~$1$ in Figure~\ref{fig:PhotoproductionStartTopologies}, is shown. The implied cosines and sines of relative phases are indicated as well.}
\label{fig:MoravcsikExample}
\end{figure}

For single-meson photoproduction, $N = 4$ amplitudes are accompanied by $N^2 = 16$ observables~\cite{Chiang:1996em} (definitions for the observables are collected in Table~\ref{tab:PhotoObservables}). The~$16$ observables can be subdivided into~$4$ \textit{shape-classes}: one class contains the four 'diagonal' observables~$\left\{ \sigma_{0}, \check{\Sigma}, \check{T}, \check{P} \right\}$, which uniquely fix the moduli~$\left| b_{1} \right|, \ldots,  \left| b_{4} \right| $, and furthermore there exist the three non-diagonal shape-classes of beam-target ($\mathcal{BT}$), beam-recoil ($\mathcal{BR}$) and target-recoil ($\mathcal{TR}$-) observables, which show a very similar mathematical structure. In order to apply Moravcsik's theorem~(sec.~\ref{sec:MoravcsiksTheorem}), the latter three shape-classes need to be further decoupled by defining~\cite{Wunderlich:2020umg}:
\begin{equation}
  \tilde{\Ocal}^{n}_{1 \pm} := \frac{1}{2} \left( \Ocal^{n}_{1+} \pm \Ocal^{n}_{1-}  \right) \text{, } \tilde{\Ocal}^{n}_{2 \pm} := \frac{1}{2} \left( \Ocal^{n}_{2+} \pm \Ocal^{n}_{2-} \right)  , \label{eq:TildeObsDefs} 
  \end{equation}
for~$n = a,b,c$. One can isolate, for example: $ \mathrm{Im} \left[ b_{4}^{\ast} b_{2}\right] = \left| b_{2} \right| \left| b_{4} \right| \sin \phi_{24} = \tilde{\Ocal}^{a}_{1-} = \frac{1}{2} \left( \Ocal^{a}_{1+} - \Ocal^{a}_{1-}  \right) = \frac{1}{2} \left(- \check{G} - \check{F} \right)$.
For $N = 4$, one has to consider~$\frac{(N - 1)!}{2} = 3$ connected graph-topologies (cf.~Figure~\ref{fig:PhotoproductionStartTopologies}). One example for a fully complete graph derived from the first topology is shown in Figure~\ref{fig:MoravcsikExample}, where the corresponding sines and cosines of relative phases are shown as well. Combining the definitions~\eqref{eq:TildeObsDefs} with Table~\ref{tab:PhotoObservables}, one can see quickly that the graph in Figure~\ref{fig:MoravcsikExample} corresponds to the observable-set~$\left\{ \check{E}, \check{H}, \check{L}_{x'}, \check{T}_{z'}, \check{L}_{z'}, \check{T}_{x'} \right\}$, which forms

\begin{table*}[ht]
 \begin{center}
 \begin{tabular}{lcr}
 \hline
 \hline
  Observable & \hspace*{5pt} Relative-phases  & \hspace*{10pt} Shape-class \\
  \hline 
  $\sigma_{0} = \frac{1}{2} \left( \left| b_{1} \right|^{2} + \left| b_{2} \right|^{2} + \left| b_{3} \right|^{2} + \left| b_{4} \right|^{2} \right)$ &  &     \\
  $- \check{\Sigma} = \frac{1}{2} \left( \left| b_{1} \right|^{2} + \left| b_{2} \right|^{2} - \left| b_{3} \right|^{2} - \left| b_{4} \right|^{2} \right)$  &  &   $\mathcal{S} = \mathrm{D}$ \\
  $- \check{T} = \frac{1}{2} \left( - \left| b_{1} \right|^{2} + \left| b_{2} \right|^{2} + \left| b_{3} \right|^{2} - \left| b_{4} \right|^{2} \right)$  &  &    \\
  $\check{P} = \frac{1}{2} \left( - \left| b_{1} \right|^{2} + \left| b_{2} \right|^{2} - \left| b_{3} \right|^{2} + \left| b_{4} \right|^{2} \right)$  &   &    \\
  \hline
   $\Ocal^{a}_{1+} = \left| b_{1} \right| \left| b_{3} \right| \sin \phi_{13} + \left| b_{2} \right| \left| b_{4} \right| \sin \phi_{24} = \mathrm{Im} \left[ b_{3}^{\ast} b_{1} + b_{4}^{\ast} b_{2} \right] = - \check{G}$  &  &  \\
   $\Ocal^{a}_{1-} = \left| b_{1} \right| \left| b_{3} \right| \sin \phi_{13} - \left| b_{2} \right| \left| b_{4} \right| \sin \phi_{24}  = \mathrm{Im} \left[ b_{3}^{\ast} b_{1} - b_{4}^{\ast} b_{2} \right] = \check{F}$  &  $\left\{ \phi_{13}, \phi_{24} \right\}$  & $a = \mathcal{BT} = \mathrm{PR}$ \\
   $\Ocal^{a}_{2+} = \left| b_{1} \right| \left| b_{3} \right| \cos \phi_{13} + \left| b_{2} \right| \left| b_{4} \right| \cos \phi_{24}  = \mathrm{Re} \left[ b_{3}^{\ast} b_{1} + b_{4}^{\ast} b_{2} \right] = - \check{E}$  &   &  \\
   $\Ocal^{a}_{2-} = \left| b_{1} \right| \left| b_{3} \right| \cos \phi_{13} - \left| b_{2} \right| \left| b_{4} \right| \cos \phi_{24} = \mathrm{Re} \left[ b_{3}^{\ast} b_{1} - b_{4}^{\ast} b_{2} \right] =  \check{H}$  &   &  \\
   \hline
   $\Ocal^{b}_{1+} = \left| b_{1} \right| \left| b_{4} \right| \sin \phi_{14} + \left| b_{2} \right| \left| b_{3} \right| \sin \phi_{23} = \mathrm{Im} \left[ b_{4}^{\ast} b_{1} + b_{3}^{\ast} b_{2} \right] = \check{O}_{z'}$  &   &  \\
   $\Ocal^{b}_{1-} = \left| b_{1} \right| \left| b_{4} \right| \sin \phi_{14} - \left| b_{2} \right| \left| b_{3} \right| \sin \phi_{23}  = \mathrm{Im} \left[ b_{4}^{\ast} b_{1} - b_{3}^{\ast} b_{2} \right] = - \check{C}_{x'}$  &  $\left\{ \phi_{14}, \phi_{23} \right\}$  & $b = \mathcal{BR} = \mathrm{AD}$ \\
   $\Ocal^{b}_{2+} = \left| b_{1} \right| \left| b_{4} \right| \cos \phi_{14} + \left| b_{2} \right| \left| b_{3} \right| \cos \phi_{23}  = \mathrm{Re} \left[ b_{4}^{\ast} b_{1} + b_{3}^{\ast} b_{2} \right] = - \check{C}_{z'}$  &   &  \\
   $\Ocal^{b}_{2-} = \left| b_{1} \right| \left| b_{4} \right| \cos \phi_{14} - \left| b_{2} \right| \left| b_{3} \right| \cos \phi_{23} = \mathrm{Re} \left[ b_{4}^{\ast} b_{1} - b_{3}^{\ast} b_{2} \right] = - \check{O}_{x'}$  &   &  \\
   \hline
   $\Ocal^{c}_{1+} = \left| b_{1} \right| \left| b_{2} \right| \sin \phi_{12} + \left| b_{3} \right| \left| b_{4} \right| \sin \phi_{34} = \mathrm{Im} \left[ b_{2}^{\ast} b_{1} + b_{4}^{\ast} b_{3} \right] = - \check{L}_{x'}$  &   &  \\
   $\Ocal^{c}_{1-} = \left| b_{1} \right| \left| b_{2} \right| \sin \phi_{12} - \left| b_{3} \right| \left| b_{4} \right| \sin \phi_{34}  = \mathrm{Im} \left[ b_{2}^{\ast} b_{1} - b_{4}^{\ast} b_{3} \right] = - \check{T}_{z'}$  &  $\left\{ \phi_{12}, \phi_{34} \right\}$  & $c = \mathcal{TR} = \mathrm{PL}$ \\
   $\Ocal^{c}_{2+} = \left| b_{1} \right| \left| b_{2} \right| \cos \phi_{12} + \left| b_{3} \right| \left| b_{4} \right| \cos \phi_{34}  = \mathrm{Re} \left[ b_{2}^{\ast} b_{1} + b_{4}^{\ast} b_{3} \right] = - \check{L}_{z'}$  &   &  \\
   $\Ocal^{c}_{2-} = \left| b_{1} \right| \left| b_{2} \right| \cos \phi_{12} - \left| b_{3} \right| \left| b_{4} \right| \cos \phi_{34} = \mathrm{Re} \left[ b_{2}^{\ast} b_{1} - b_{4}^{\ast} b_{3} \right] = \check{T}_{x'}$   &   &  \\
   \hline
   \hline
 \end{tabular}
 \end{center}
 \caption{The $16$ polarization ob\-serva\-bles in pseudoscalar meson photoproduction (cf. ref.~\cite{Chiang:1996em}) are written here in terms of transversity-amplitudes~$b_{1}, \ldots, b_{4}$. The non-diagonal observables are given in the symbolic notation $\Ocal^{n}_{\nu \pm}$ by Nakayama (cf. reference~\cite{Nakayama:2018yzw}), as well as in the more commonly used notation (rightmost names). The subdivision into $4$ shape-classes, with corresponding relative-phases, is indicated.}
 \label{tab:PhotoObservables}
\end{table*}

\noindent
a complete set of~$10$ when combined with the four 'diagonal' observables. Investigating further all possible graphs with an odd number of dashed lines derived from each of the~$3$ start-topologies,~$12$ complete sets composed of~$10$ observables were found in ref.~\cite{Wunderlich:2020umg}, each containing two observables more than the absolutely minimal number of~$2 N = 8$. 

For electroproduction~($N = 6$)~\cite{Tiator:2017cde}, the~$N^{2} = 36$ observables can be subdivided into~$10$ shape-classes and~$6$ 'diagonal' observables fix the moduli~$\left| b_{1} \right|, \ldots,  \left| b_{6} \right|$. When applying Moravcsik's theorem~(sec.~\ref{sec:MoravcsiksTheorem}) to each of the~$\frac{(N - 1)!}{2} = 60$ relevant graph-topologies~(see figures in ref.~\cite{Wunderlich:2020umg}), one obtains~$776$ complete sets in total, with a minimum length of~$13$ observables~\cite{Wunderlich:2020umg}. This is still one observable more than the absolutely minimal number of~$2 N = 12$. 

\section{New 'directional' graphs} \label{sec:NewDirectionalGraphs}

\begin{figure}[ht]
\centering
\includegraphics[width=0.825\textwidth,trim = 0 5.5cm 0 1.15cm,clip]{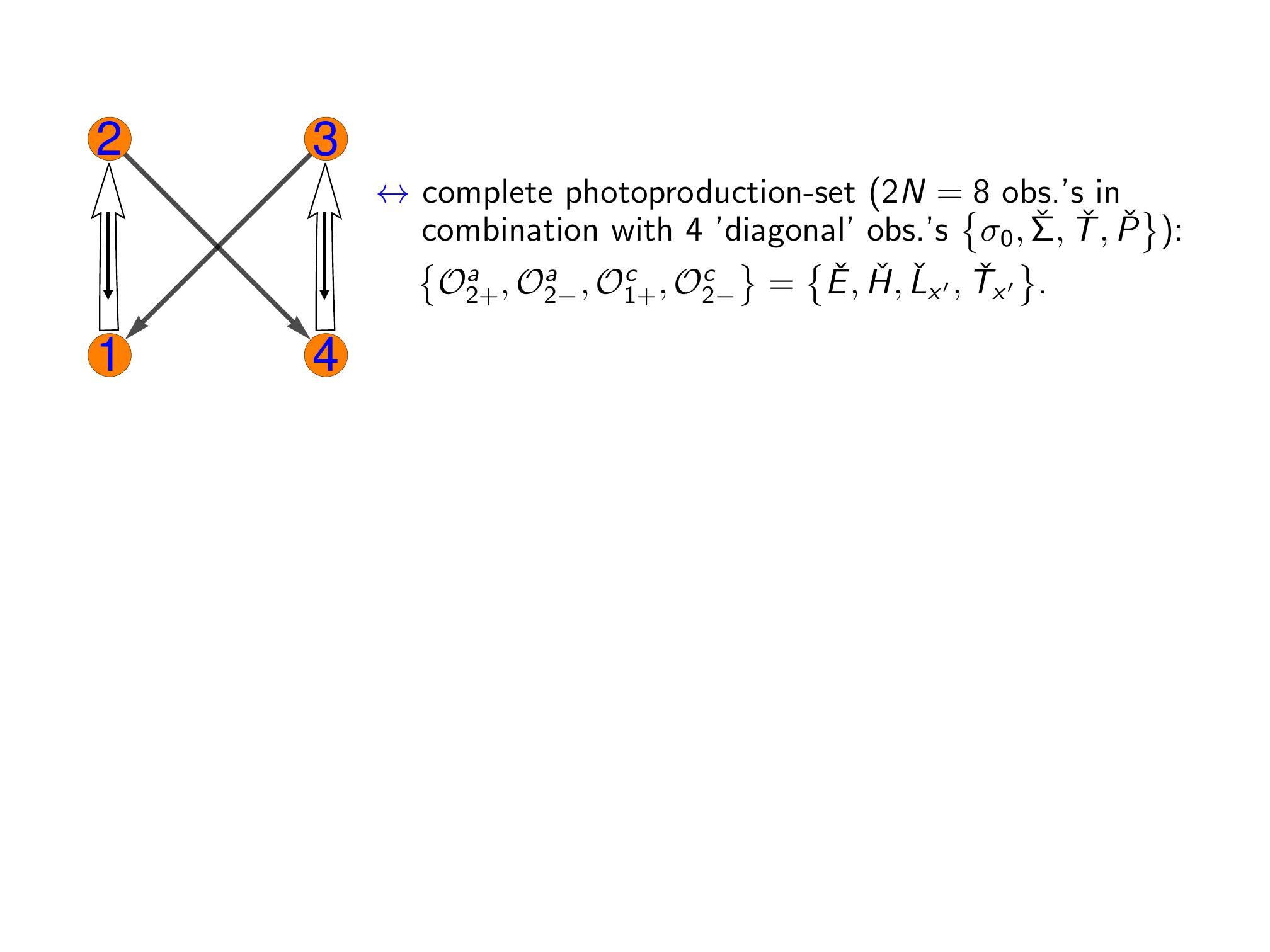} 
\vspace*{-12.5pt}
\caption{One example for the fully complete graphs according to the new criterion using graphs with directional information (cf. section~\ref{sec:NewDirectionalGraphs} and reference~\cite{Wunderlich:2021xhp}), which can be derived for single-meson photoproduction ($N = 4$ amplitudes), is depicted. The implied complete set with $2 N = 8$ observables is indicated as well.}
\label{fig:NewGraphExample}
\end{figure}

In order to reduce the mismatch between the results of Moravcsik's theorem and the absolutely minimal length of~$2N$ observables in a complete set (cf. sec~\ref{sec:ApplicationsPhotoElectroproduction}), new graphs have been proposed in reference~\cite{Wunderlich:2021xhp}, which contain additional directional information. The new completeness criterion for these graphs makes heavy use of the repeating mathematical structure of non-diagonal 'shape-classes of~$4$' (cf.~Table~\ref{tab:PhotoObservables}). One example for a 'directional' graph for photoproduction~($N = 4$), as well as a corresponding complete set, is shown in Figure~\ref{fig:NewGraphExample}. The new graphs contain~\cite{Wunderlich:2021xhp}:
\begin{itemize}
      \item[-] Single-lined arrows, which have the same meaning as in Moravcsik's theorem (sec.~\ref{sec:MoravcsiksTheorem}). 
      \item[-] Double-lined arrows denoting 'crossed' selections of observable-pairs, i.e. $\Ocal^{n}_{1 \pm}$ $\oplus$ $\Ocal^{n}_{2 \pm}$ (e.g.:~$\left( \Ocal^{n}_{1 +}, \Ocal^{n}_{2 +} \right)$). 
      \item[-] An 'outer' direction which is in direct correspondence to a directional convention for indices appearing in the consistency relation.  For the graph in Figure~\ref{fig:NewGraphExample}, the consistency relation reads: $\phi_{12} + \phi_{24} + \phi_{43} + \phi_{31}  = 0$.
      \item[-] The direction of smaller 'inner' arrows inside of double-lined arrows, which specifies the sign of '$\zeta$-angles' appearing in the formulas for discrete phase-ambiguities~(see~\cite{Wunderlich:2021xhp}). The $\zeta$-angles are important for removing dependencies among consistency relations.
      \item[$\Rightarrow$] The graph in Figure~\ref{fig:NewGraphExample}, for example, is complete because the 'inner' arrows both point opposite to the bigger double-lined arrows (cf. explanations in ref.~\cite{Wunderlich:2021xhp}).
 \end{itemize}

Using these new graphs, $60$ known complete sets of minimal length~$2N$ have been re-derived for photoproduction and $1216$ new such sets have been found for electroproduction (see ref.~\cite{Wunderlich:2021xhp}). Reactions with~$N > 6$ amplitudes likely require new and more involved algebraic derivations~\cite{Wunderlich:2021xhp}.

\section{Problems with $N > 6$ amplitudes} \label{sec:HigherNCases}

The graphical methods discussed up to this point show their real utility once applied to problems with larger numbers of amplitudes, for instance~$N > 6$. The connectedness of the graphs automatically removes continuous ambiguities and it ensures the possibility to write down a consistency relation, two tasks which can become rather awkward when done 'by hand' for problems with higher~$N$.

\clearpage

\begin{figure}[h]
\centering
\includegraphics[width=0.299\textwidth,trim = 0 0 0 0.1cm,clip]{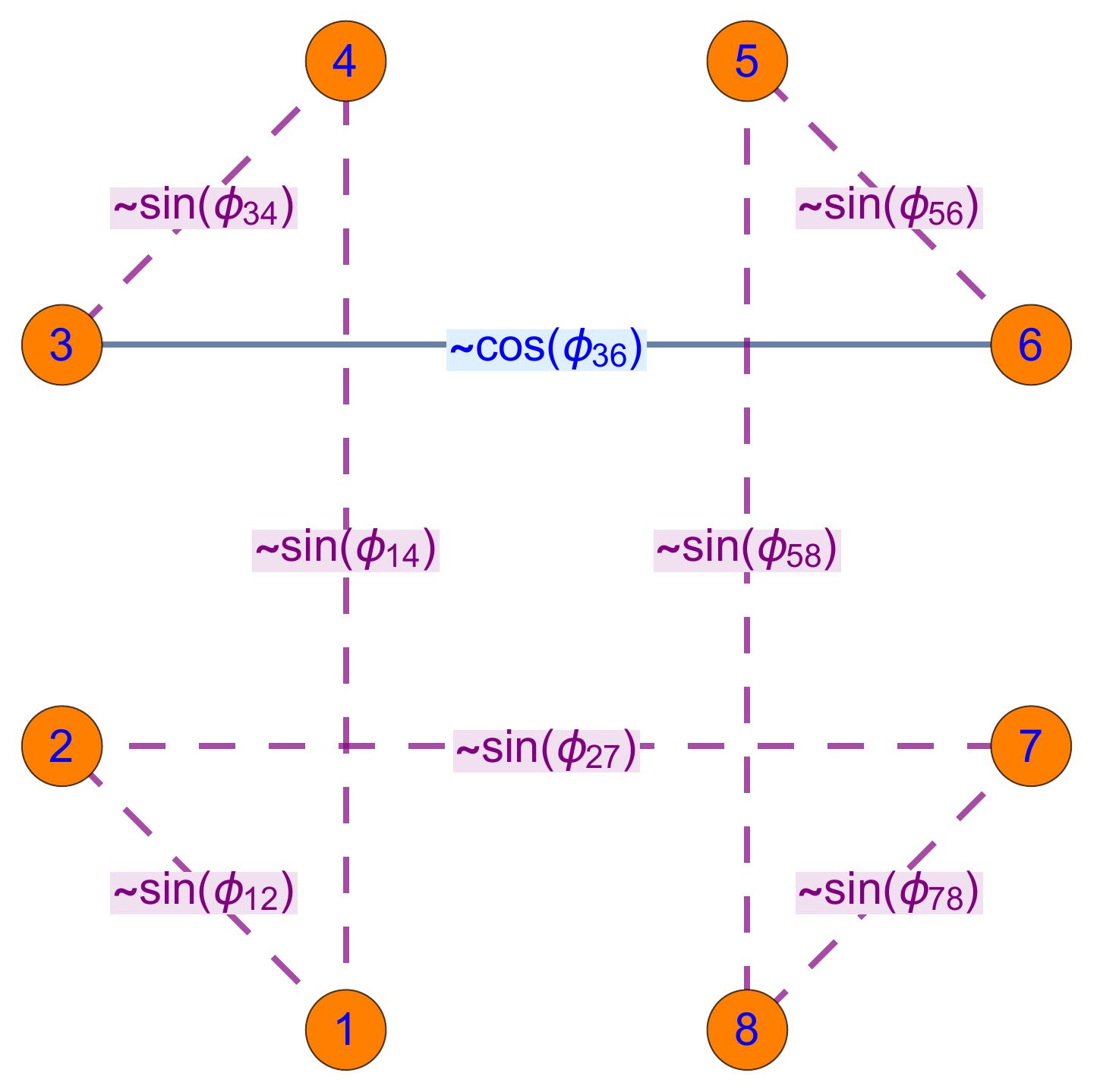} \hspace*{30pt}
\includegraphics[width=0.299\textwidth,trim = 0 0 0 0.1cm,clip]{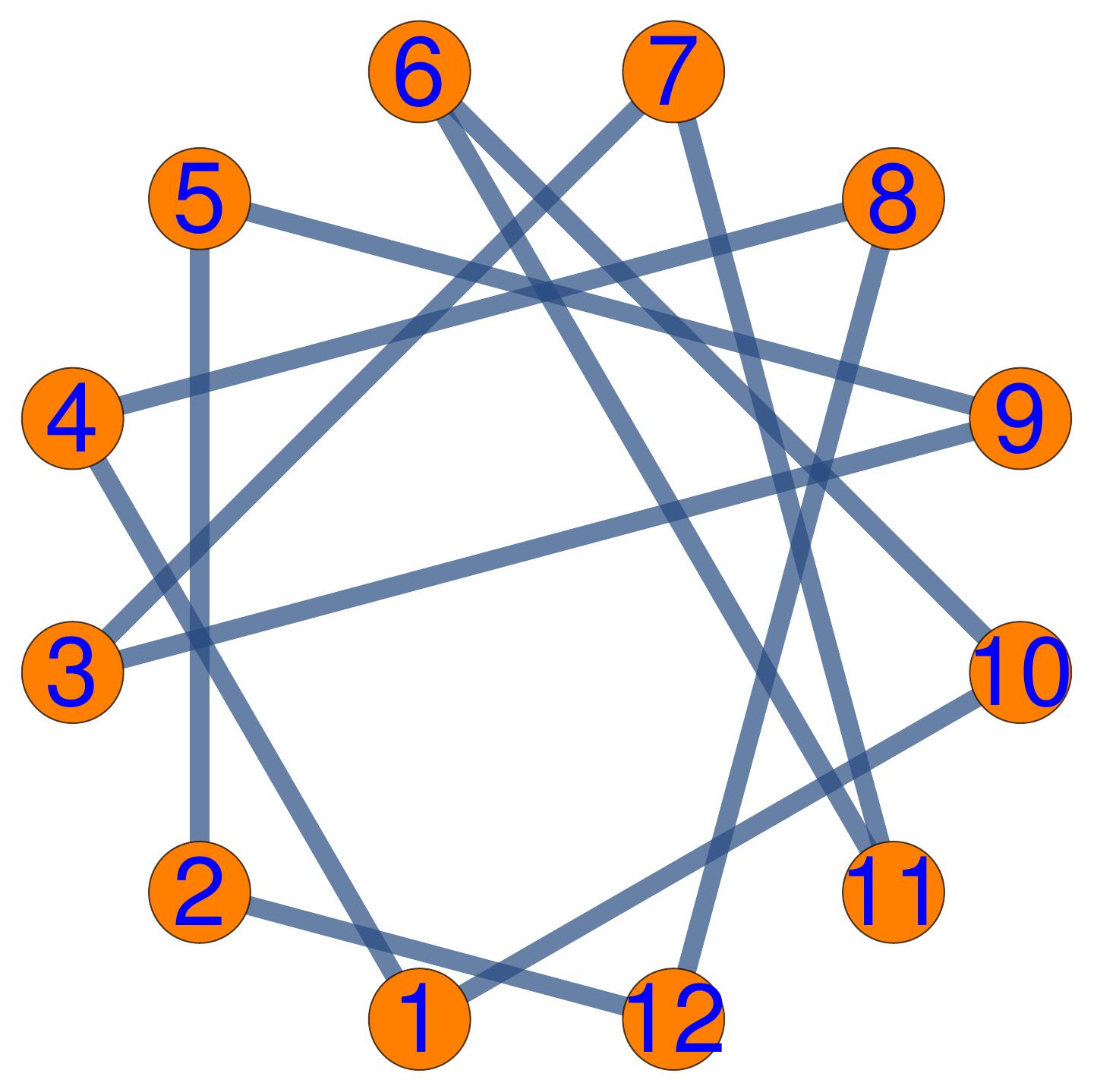}
\vspace*{-7.5pt}
\caption{Left: An example-graph for two-meson photoproduction ($N = 8$ amplitudes) is shown (figure taken from reference~\cite{Kroenert:2020ahf}), which is fully complete according to the modified form of Moravcsik's theorem (cf. section~\ref{sec:MoravcsiksTheorem}). Right: A fully connected example-topology is shown for the problem of vector-meson photoproduction ($N = 12$ amplitudes).}
\label{fig:EightAndTwelveAmplitudesExampleGraphs}
\end{figure}

For two-meson photoproduction~($N = 8$)~\cite{Roberts:2004mn}, the selection of complete sets from the full set of~$N^{2} = 64$ observables proceeds via~$\frac{(N - 1)!}{2} = 2520$ connected start-topologies. One example for a fully complete graph according to Moravcsik~(sec.~\ref{sec:MoravcsiksTheorem}) is shown in Figure~\ref{fig:EightAndTwelveAmplitudesExampleGraphs}, on the left. In reference~\cite{Kroenert:2020ahf}, it was found that the Moravcsik-complete sets for this problem contain at least~$24$ observables. Numerical as well as algebraic methods were than applied to reduce these sets to the minimal length of~$2 N = 16$. For more details on applications to two-meson photoproduction, see the contribution~\cite{Kroenert:HADRON2021} to this conference.

For vector-meson photoproduction~($N = 12$)~\cite{Pichowsky:1994gh}, one examplary connected start-topology is shown in Figure~\ref{fig:EightAndTwelveAmplitudesExampleGraphs}, on the right. The graphical solution to this problem can be automated on a computer, even though the number of~$\frac{(N - 1)!}{2} = 19958400$ connected start-topologies may look a bit in\-timi\-dat\-ing. Vector-meson photoproduction is a possible future option for applications of the methods presented here. 

\section{Conclusions and Outlook} \label{sec:Conclusions}

Graphical methods for the determination of complete experiments have been discussed. These methods have reproduced already known results for reactions with relatively few spin amplitudes, e.g. single-meson photoproduction~($N = 4$ amplitudes). The real power of such graphical methods lies in their capability to derive complete sets for more complicated reactions, i.e. with higher numbers of~$N > 6$ spin amplitudes, where these methods should be applied in order to help plan future measurements in baryon spectroscopy.


\end{multicols}


\vspace*{-7.5pt}

\vspace*{-7.5pt}

\vspace*{-5pt}

\clearpage

\vspace*{-20pt}

\vspace*{-17.5pt}

\vspace*{-16.5pt}

\clearpage

\listoftables

\listoffigures

\clearpage

\medline
\begin{multicols}{2}

\end{multicols}
\end{document}